\documentclass[twocolumn,showpacs,preprintnumbers,amsmath,amssymb]{revtex4-1}

\usepackage{graphicx}
\usepackage{dcolumn}
\usepackage{simplewick}
\usepackage{bm}
\usepackage{hyperref}
\usepackage{dcolumn}
\usepackage[usenames,dvipsnames]{color}

\def\red{|\!|}
\def\bfd{\mathbf{D}}
\def\pt{\mathbf{T}^{(1)}}
\def\pto{\mathbf{T}^{(1)}_1}
\def\ptt{\mathbf{T}^{(1)}_2}
\def\ps{\mathbf{S}^{(1)}}
\def\pso{\mathbf{S}^{(1)}_1}
\def\pst{\mathbf{S}^{(1)}_2}
\def\helec{{\mathbf{H}}_{\rm elec}^{\rm NSD}}

\begin{document}

\title{Relativistic coupled-cluster calculations of nuclear spin-dependent
       parity non-conservation in Cs, Ba$^+$ and Ra$^+$}

\author{B. K. Mani and D. Angom}
\affiliation{Physical Research Laboratory,
             Navarangpura-380009, Gujarat,
             India}

\begin{abstract}
  We have developed a relativistic coupled-cluster theory to incorporate 
nuclear spin-dependent interaction Hamiltonians perturbatively. This theory
is ideal to calculate parity violating nuclear spin-dependent electric 
dipole transition amplitudes, $E1_{\rm PNC}^{\rm NSD}$, of heavy atoms. 
Experimental observation of which is a clear signature of 
nuclear anapole moment, the dominant source of nuclear spin-dependent parity
violation in atoms and ions. We apply the theory to calculate 
$E1_{\rm PNC}^{\rm NSD}$ of Cs, which to date has provided the best atomic
parity violation measurements. We also calculate $E1_{\rm PNC}^{\rm NSD}$ of
Ba$^+ $ and Ra$^+$, candidates of ongoing and proposed experiments. 
\end{abstract}



\maketitle


  The effects of parity nonconservation (PNC) in atoms occur in two forms,
nuclear spin-independent (NSI) and nuclear spin-dependent (NSD). The former
is well studied and experimentally observed in several atoms. The signature
of the later (NSD) has been observed only in one experiment with
Cs \cite{wood-97} and the same experiment has provided the most accurate 
results on NSI atomic PNC as well. In an atom or ion the most dominant source 
of NSD-PNC is the nuclear anapole moment (NAM), a parity odd nuclear 
electromagnetic moment. It was first suggested by Zeldovich 
\cite{zeldovich-58} and arises from parity violating phenomena within the 
nucleus.

 One major hurdle to a clear and unambiguous observation of NAM is the large 
NSI signal, which overwhelms the NSD signal. However, proposed experiments 
with single Ba$^+$ ion \cite{fortson-93} could probe PNC in the 
$s_{1/2}-d_{5/2}$ transition, where the NSI component is zero. This could then 
provide an unambiguous observation of NSD-PNC and NAM in particular. The 
ongoing experiments with atomic Ytterbium \cite{tsigutkin-09} is another 
possibility, the $6s^2\; ^1S_0-6s5d\; ^3D_2 $ transition, to observe NSD-PNC 
with minimal mixture from the NSI component. One crucial input, which is also 
the source of large uncertainty, to extract the value of NAM is the input from 
atomic theory calculations. Considering this, it is important to employ 
reliable and accurate many-body theory in the atomic theory calculations. 

    The coupled-cluster (CC) theory\cite{coester-58,coester-60} 
is one of the most reliable many-body theory to incorporate electron 
correlation in atomic calculations.  It has been used with great success in 
nuclear \cite{hagen-08}, atomic \cite{eliav-94,nataraj-08,pal-07}, molecular
\cite{isaev-04} and condensed matter \cite{bishop-09} physics. 
In atomic physics, the relativistic coupled-cluster (RCC) theory has been used 
extensively in atomic properties calculations, for example, hyperfine 
structure constants \cite{pal-07,sahoo-09} and electromagnetic 
transition properties \cite{thierfelder-09,sahoo-09a}. In atomic PNC 
calculations too, RCC is the preferred theory and several groups have
used it to calculate NSI-PNC of atoms \cite{wansbeek-08,pal-09,porsev-10}. 
However, the calculations in Ref. \cite{wansbeek-08} are entirely based on
RCC with a variation we refer to as perturbed RCC (PRCC), where as 
the calculations in Ref. \cite{pal-09,porsev-10} are based on sum over states 
with CC wave functions. 

  To date, the use of PRCC in atomic PNC
is limited to NSI-PNC. In this letter we report the PRCC theory to calculate
NSD-PNC in atoms. Such a development is timely as the recent experimental 
proposals on Ba$^+ $ and Ra$^+$ \cite{sahoo-11} and observation of large
enhancement in atomic Yb \cite{tsigutkin-09} shall require precision atomic 
theory to examine the systematics and interpret the results. It must perhaps 
be mentioned that, in an earlier work we had developed and calculated electric 
dipole moment of atomic Hg \cite{latha-09} using PRCC theory.


{\em RCC theory}.---In the RCC method, the atomic state is expressed in terms 
of $T$  and $S$, the closed-shell and one-valence cluster operators 
respectively, as
\begin{equation}
  |\Psi_v\rangle = e^{T^{(0)}} \left [  1 + S^{(0)} \right ] |\Phi_v\rangle,
  \label{cceqn_1v}
\end{equation}
where $|\Phi_v\rangle$ is the one-valence Dirac-Fock reference state. It is 
obtained by adding an electron to the closed-shell reference state,
$|\Phi_v \rangle = a^\dagger_v|\Phi_0\rangle$. In the coupled-cluster singles 
doubles (CCSD) approximation $T^{(0)} = T^{(0)}_1 + T^{(0)}_2$ and 
$S^{(0)} = S^{(0)}_1 + S^{(0)}_2$. The open-shell cluster operators are 
solutions of the nonlinear equations \cite{mani-10}
\begin{subequations}
\label{cc_sin_dou}
\begin{eqnarray}
  \langle \Phi_v^p|\bar H_N \! +\! \{\contraction[0.5ex]
  {\bar}{H}{_N}{S} \bar H_N S^{(0)}\} |\Phi_v\rangle
  &=&E_v^{\rm att}\langle\Phi_v^p|S^{(0)}_1|\Phi_v\rangle ,
  \label{ccsingles}     \\
  \langle \Phi_{va}^{pq}|\bar H_N +\{\contraction[0.5ex]
  {\bar}{H}{_N}{S}\bar H_N S^{(0)}\} |\Phi_v\rangle
  &=& E_v^{\rm att}\langle\Phi_{va}^{pq}|S^{(0)}_2|\Phi_v\rangle,
  \label{ccdoubles}
\end{eqnarray}
\end{subequations}
where $\bar H_{\rm N}=e^{-T^{(0)}}H_{\rm N}e^{T^{(0)}} $ is the
similarity transformed Hamiltonian and  the normal ordered atomic  Hamiltonian
$H_{\rm N} = H -\langle\Phi_0|H|\Phi_0\rangle$. And, 
$E_v^{\rm att} = E_v - E_0,$ is the attachment energy of the
valence electron. The $T^{(0)}$ are solutions of a similar set of equations,
however, with $ S^{(0)}=0$. A similar set of equations may be derived in the 
case of two-valence systems and use it in the wave function and properties 
calculations of atoms like Yb \cite{mani-11}.

\begin{figure}
\begin{center}
  \includegraphics[width = 7.0 cm]{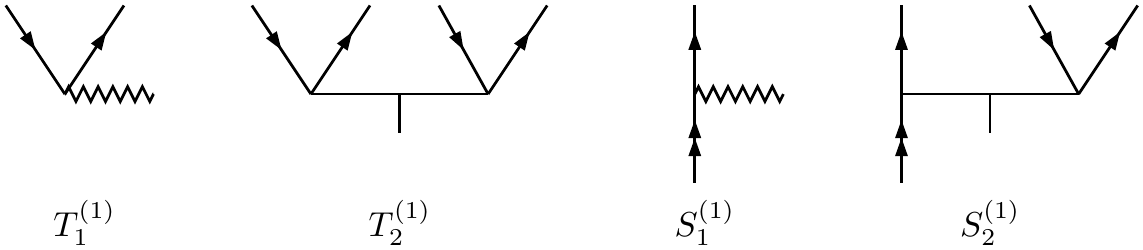}
  \caption{Diagrammatic representation of single and double excitation 
           perturbed cluster operators. The short line on the interaction  
           line of $\ptt$ and $\pst$ is to indicate the multipole structure
           of these operators.}
  \label{pert_cc_op}
\end{center}
\end{figure}

{\em Perturbed RCC theory}.---The perturbed RCC method \cite{sahoo-06,mani-09},
unlike the standard time-independent perturbation theory, implicitly accounts 
for all the possible intermediate states in properties calculations. Consider 
the NSD-PNC interaction Hamiltonian
\begin{equation}
   H_{\rm PNC}^{\rm NSD}=\frac{G_{\rm F}\mu'_W}{\sqrt{2}}\sum_i
   \bm{\alpha}_i\cdot \mathbf{I}\rho_{\rm{N}}(r),
  \label{hpncnsd2}
\end{equation}
as the perturbation. Here, $\mu'_W$ is the weak nuclear moment of the nucleus 
and $\rho_{\rm N}(r)$ is the nuclear density. The total atomic Hamiltonian is 
\begin{equation}
    H_{\rm A} = H^{\rm DC} + \lambda H_{\rm PNC}^{\rm NSD},
    \label{total_H}
\end{equation}
where $\lambda$ is the perturbation parameter. Mixed parity hyperfine states 
$|\widetilde{\Psi}_v \rangle$ are then the eigen states of $H_{\rm A}$.
To calculate $|\widetilde{\Psi}_v \rangle$ from RCC,  we define a new set of 
cluster operators $\pt$, which unlike $T^{(0)}$ connects the reference state to 
opposite parity states. This is the result of incorporating one order of 
$H_{\rm PNC}^{\rm NSD}$ and for this reason we refer to $\pt $ as the 
perturbed cluster operators. Although hyperfine states are natural to
$H_{\rm PNC}^{\rm NSD}$,  cluster operator $\pt $ is defined to operate only 
in the electronic space and is a rank one operator. For this define 
$\helec = (G_{\rm F}\mu'_W)/(\sqrt{2})\sum_i
\bm{\alpha}_i\rho_{\rm{N}}(r)$, which operates only in the electronic space,  
so that $ H_{\rm PNC}^{\rm NSD}=  \helec\cdot \mathbf{I}$.  The closed-shell 
exponential operator in PRCC is $ e^{T^{(0)}+ \lambda\pt\cdot\mathbf{I}}$ and 
the atomic state is 
\begin{equation}
  |\widetilde \Psi_0 \rangle = e^{T^{(0)}} \left [  1 
    + \lambda \pt\cdot\mathbf{I} \right ] |\Phi_0\rangle.
  \label{pccwave2}
\end{equation}
Similarly, the mixed parity state from one-valence PRCC theory is 
\begin{equation}
  | \widetilde{\Psi}_v \rangle = e^{T^{(0)}}\left[ 1 
    + \lambda \pt \cdot\mathbf{I} \right] \left[ 1
    + S^{(0)} +\lambda \ps \cdot\mathbf{I} \right] |\Phi_v \rangle.
  \label{psiptrb1v}
\end{equation}
As $\pto$ is one particle and rank one operator, in terms of c-tensors
\begin{equation}
  \pto = \sum_{ap}\tau_a^p \mathbf{C}_1(\hat{r}),
\end{equation}
where $\mathbf{C}_i $ are c-tensor operators. Similarly, the tensor structure 
of $\ptt$  is 
\begin{equation}
   \ptt = \sum_{abpq}\sum_{l_1, l_2} \tau_{ab}^{pq}(l_1, l_2) 
       \{\mathbf{C}_{l_1}(\hat{r}_1)\mathbf{C}_{l_2}(\hat{r}_2)  \}^1, 
\end{equation}
where $\{\cdots \}^1 $ indicates the two c-tensor operators couple to a rank 
one tensor operator. Based on the tensor structures, the perturbed cluster 
operators are diagrammatically represented as shown in Fig. \ref{pert_cc_op}. 
For the doubles $\ptt $, to indicate the multipole structure,  an additional 
line is added to the interaction line. The cluster operators are solutions
of the equations
\begin{subequations}
\begin{eqnarray}
  && \langle \Phi^p_v |\{ \contraction{}{H}{_{\rm N}}{S}\bar{H}_{\rm N}
     \mathbf{S}^{(1)} \} + \{ \contraction{}{H}{_{\rm N}}{S}\bar{H}_{\rm N}
     \mathbf{T}^{(1)} \} + \{ \contraction{}{H}{_{\rm N}}{T}
     \contraction[1.5ex]{}{V}{_{\rm N}T^{(1)}}{S}H_{\rm N}
     \mathbf{T}^{(1)}S^{(0)}\} + \bar{\mathbf{H}}_{\rm elec}^{\rm NSD} 
                   \nonumber \\
   && + \{ \contraction{}{H}{_{\rm elec}^{\rm NSD}}{S}
     \bar{\mathbf{H}}_{\rm elec}^{\rm NSD}{S}^{(0)} \}|\Phi_v \rangle =
  \Delta E_v \langle \Phi^p_v | \pso|\Phi_v \rangle,
  \label{ccsptrb1v1}
\end{eqnarray}
\begin{eqnarray}
  && \langle \Phi^{pq}_{vb} |\{ \contraction{}{H}{_{\rm N}}{S}\bar{H}_{\rm N}
     \mathbf{S}^{(1)} \} + \{ \contraction{}{H}{_{\rm N}}{S}\bar{H}_{\rm N}
     \mathbf{T}^{(1)} \} + \{ \contraction{}{H}{_{\rm N}}{T}
     \contraction[1.5ex]{}{V}{_{\rm N}T^{(1)}}{S}H_{\rm N}
     \mathbf{T}^{(1)}S^{(0)}\} + \bar{\mathbf{H}}_{\rm elec}^{\rm NSD} 
                               \nonumber \\
  &&  + \{ \contraction{}{H}{_{\rm elec}^{\rm NSD}}{S}
     \bar{\mathbf{H}}_{\rm elec}^{\rm NSD}{S}^{(0)} \}|\Phi_v \rangle =
  \Delta E_v \langle \Phi^{pq}_{vb} | \pst|\Phi_v \rangle,
  \label{ccsptrb1v2}
\end{eqnarray}
\label{prcc_eqn}
\end{subequations}
Where we have used the relations $ \langle \Phi^p_v | \pt |\Phi_v \rangle = 0$,
and $\langle \Phi^p_v| \pt S| \Phi_v \rangle = 0$, as the bra state is valence 
excited. An approximate form of Eq. (\ref{prcc_eqn}), but which contains all 
the important many-body effects, are the linearized cluster equations. This is 
obtained by considering
$   \contraction{}{H}{_{\rm N}}{T}\bar{H}_{\rm N}\mathbf{T}^{(1)} \approx
        \contraction{}{H}{_{\rm N}}{T}{H}_{\rm N}\mathbf{T}^{(1)}$, and
$   \bar {\mathbf{H}}_{\rm elec}^{\rm NSD}  \approx   
      \mathbf{H}_{\rm elec}^{\rm NSD} + 
   \contraction{}{H}{_{\rm elec}^{\rm NSD}}{T}
       {\mathbf{H}}_{\rm elec}^{\rm NSD}T^{(0)}$. We refer to this as the 
linear approximation and use it extensively to check the results.
%
%
\vspace{0.2cm}

{\em $E_{\rm PNC}^{\rm NSD}$ calculations}.---If $|\Psi_v\rangle$ and 
$|\Psi_w\rangle$ are atomic states of same parity, then the 
$H_{\rm PNC}^{\rm NSD}$ induced electric dipole transition amplitude
$E1_{\rm PNC}^{\rm NSD}=\langle \widetilde{\Psi}_w \red \mathbf{D} \red
\widetilde{\Psi}_v \rangle$, where $\mathbf{D}$ is the dipole operator.
Similarly, the transition amplitude within the electronic sector is 
\begin{eqnarray}
  E1_{\rm elec}^{\rm NSD}& = & \langle \Phi_w \red \bar{\mathbf{D}} 
     \left[ \pt + \ps + \pt S \right]+ \left[ \pt + \ps  \right. 
                                              \nonumber \\ 
              && \left. + \pt S \right]^\dagger \bar{\mathbf{D}} +
                 S^\dagger\bar{\mathbf{D}} \left[ \pt + \ps + \pt S 
                  \right] \nonumber \\
              && + \left[ \pt + \ps + \pt S \right]^\dagger
                 \bar{\mathbf{D}} S \red \Phi_v \rangle,
\end{eqnarray}
where $\bar{\mathbf{D}} = {e^T}^\dagger \bm{\mathbf{D}} e^T,$ is the dressed 
electric dipole operator. It is evident that $\bar{\mathbf{D}}$ is a 
non-terminating series of the closed-shell cluster operators. It is 
non-trivial to incorporate $T$  to all orders in numerical computations. For 
this reason $\bar{\mathbf{D}}$ approximated as
$ \bar{\mathbf{D}} \approx \mathbf{D} + \mathbf{D} T^{(0)} + 
  {T^{(0)}}^\dagger \mathbf{D} + {T^{(0)}}^\dagger \mathbf{D} T^{(0)}$.
This captures all the important contributions arising from the 
core-polarization and pair-correlation effects. Terms not included in 
this approximation are third and higher order in $T^{(0)}$.
The  expression used in our calculations is then
\begin{eqnarray}
   E1_{\rm elec}^{\rm NSD}&  \approx & \langle\Phi_w\red 
               \mathbf{D} \pt + {T^{(0)}}^\dagger \mathbf{D} \pt + 
               {\pt}^\dagger \mathbf{D} T^{(0)}  
                                                 \nonumber \\
            && +{\pt}^\dagger \mathbf{D} + \mathbf{D} \pt S^{(0)} 
               + {\pt}^\dagger {S^{(0)}}^\dagger \mathbf{D} 
                                                 \nonumber \\
            && +{S^{(0)}}^\dagger \mathbf{D} \pt
               + {\pt}^\dagger \mathbf{D} S^{(0)} + \mathbf{D} \ps
               +{\ps}^\dagger \mathbf{D}                      \nonumber \\
            && + {S^{(0)}}^\dagger \mathbf{D} 
               \ps + {\ps}^\dagger \mathbf{D} S^{(0)} \red\Phi_v\rangle.
  \label{e1pnc_elec}
\end{eqnarray}
From our previous study of properties calculations 
\cite{mani-10}, we conclude that the contributions from the higher order 
are negligible.

\begin{figure}[h]
\begin{center}
  \includegraphics[width = 6.5cm]{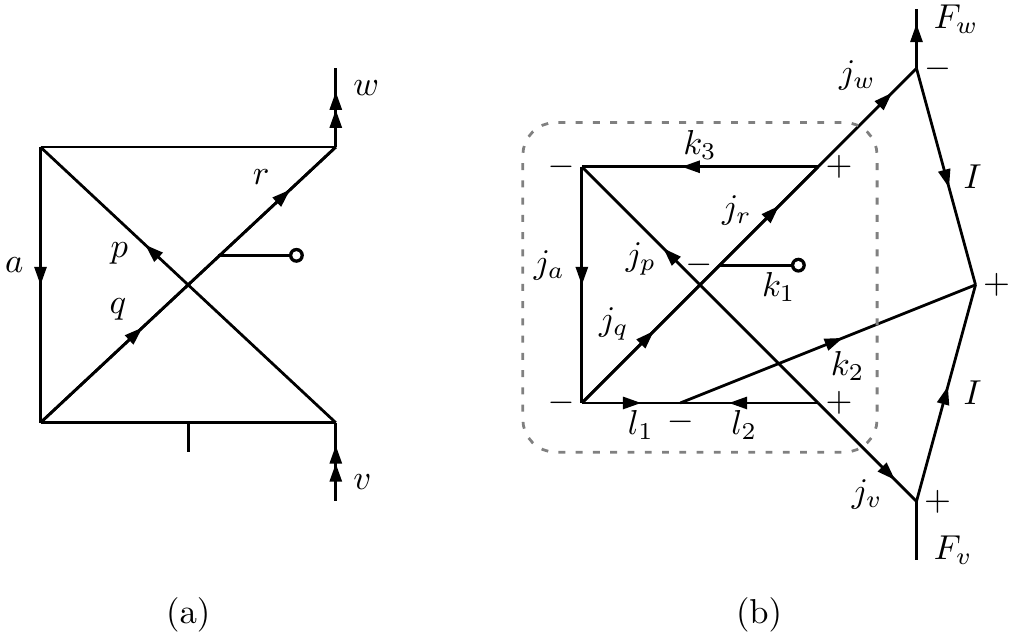}
  \caption{Examples of $E1_{\rm PNC}^{\rm NSD}$ diagrams (a) one of the 
           exchange diagrams in electronic sector and (b) angular momentum 
           diagram in terms of hyperfine states and the portion within the 
           dash lines is the electronic component.}
  \label{e1pnc_d2}
\end{center}
\end{figure}

{\em Coupling with nuclear spin}.---To couple $E1_{\rm elec}^{\rm NSD}$  with 
nuclear spin $\mathbf{I}$  and obtain $E1_{\rm PNC}^{\rm NSD}$, consider the 
exchange diagram in Fig. \ref{e1pnc_d2}(a). It arises from the term 
${T_2^{(0)}}^{\dagger}D\pst$ in the PRCC expression of 
$E_{\rm elec}^{\rm NSD}$. 

 To demonstrate the non-trivial angular integration, in hyperfine atomic 
states, the angular momentum diagram of the same diagram is shown in 
Fig. \ref{e1pnc_d2}(b). Conventions of phase and angular momentum lines of
Lindgren and Morrison \cite{lindgren-85} are used while drawing the 
diagram. The 
portion of the diagram within the rectangle in dashed-line is the angular 
momentum part of the electronic sector. 
The evaluation of the angular integral of the electronic sector, 
following Wigner-Eckert theorem, is equivalent to 
\begin{eqnarray}
   && \langle j_wm_w| \sum_{l}\left \{ {T_2^{(0)}}^{\dagger}\bfd\pso 
        \right\}^{l} |j_vm_v\rangle  = (-1)^{j_w - m_w}  \nonumber \\
  && \times \sum _l\left ( \begin{array}{ccc}
                             j_w & l & j_v \\
                            -m_w & q & m_v 
                          \end{array} \right ) \langle j_w\red
      \left \{{T_2^{(0)}}^{\dagger} \bfd\pso \right\}^{l} \red j_v\rangle,
  \label{deff_dbl_elec}
\end{eqnarray}
where $\{\ldots\}^{l}$ represents coupling of rank one tensor operators 
$\mathbf{D} $ and $\ps$ to an operator of rank $l$. This coupling is a 
structure common to any PRCC term of $E1_{\rm elec}^{\rm NSD}$. 
From the triangular condition, $l = 0, 1, 2$ are the allowed values, 
however, what values of $l$ contribute depends on $j_v$ and $j_w$. For 
example, $l=0,1$ contribute in the PNC $6\;^2S_{1/2}\rightarrow 7\;^2S_{1/2}$ 
transition of atomic Cs \cite{wood-97}, where as only $l=2$  contributes
to the proposed PNC $6\;^2S_{1/2} \rightarrow 5\; ^2D_{5/2}$  transition in 
Ba$^+$ \cite{fortson-93}.

  The angular momentum diagram in Fig. \ref{e1pnc_d2}(b), after evaluation, 
reduces to a $9j$-symbol and free line part. Algebraically, the matrix
element in the hyperfine states is
\begin{eqnarray}
   && \sum_{l}\langle F_wm_w| \left \{ \left [ {T_2^{(0)}}^{\dagger}\bfd\pso 
        \right]^{l} \mathbf{I}\right \}^1 |F_vm_v\rangle  =  (-1)^{F_w - m_w}  \nonumber \\
  && \times \left ( \begin{array}{ccc}
                       F_w & 1 & F_v \\
                      -m_w & q & m_v 
                    \end{array} \right ) \langle F_w\red
      D_{\rm eff}\red F_v\rangle,
  \label{deff_dbl}
\end{eqnarray}
where
$
  D_{\rm eff} = \sum_l\{ [ 
      {T_2^{(0)}}^{\dagger}\bfd\pso ]^{l} \mathbf{I} \}^1 , 
$
is the effective dipole operator in the hyperfine states.  As seen
from the angular momentum diagram, coupling of angular momenta in 
proper sequence is essential to obtain correct angular factors. However, the 
sequence is not manifest in the algebraic expression.

%
%
\begin{table}
\begin{center}
\caption{Reduced matrix element, $E1_{\rm PNC}^{\rm NSD}$, of the 
         $6\;^2S_{1/2} \rightarrow 7\;^2S_{1/2}$,
         $6\;^2S_{1/2} \rightarrow 5\;^2D_{3/2}$  and
         $7\;^2S_{1/2} \rightarrow 6\;^2D_{3/2}$ transitions between
         different hyperfine states in  Cs, Ba$^+$ and Ra$^+$ respectively. 
         The values listed are in units of $iea_0\times 10^{-12}\mu'_W$.}
\begin{ruledtabular}
 \label{e1pnc_table}
\begin{tabular}{ccccccc}
Atom & \multicolumn{2}{c}{Transition} &\multicolumn{3}{c}{This work} 
                                      &Other works                          \\
\hline
  & $F_f$    &    $F_i$    &    DF    &    MBPT  &  PRCC   &                \\
\hline
$^{133}$Cs&  
     3 & 3  & $2.011$ & $2.060$ & $2.274$ & $2.249$ \cite{johnson-03}       \\
  &  4 & 4  & $2.289$ & $2.338$ & $2.589$ & $2.560$ \cite{johnson-03}       \\
  &  4 & 3  & $5.000$ & $4.819$ & $5.446$ & $6.432$ \cite{johnson-03},      \\
  &    &    &         &         &         & $7.057$ \cite{safronova-09}     \\
  &  3 & 4  & $5.774$ & $5.662$ & $6.313$ & $7.299$ \cite{johnson-03},      \\
  &    &    &         &         &         & $7.948$ \cite{safronova-09}     \\
\hline
$^{135}$Ba$^+$&  
     3 & 2  & $-2.716$ & $-2.881$ & $-2.404$ & $-2.915$ \cite{sahoo-11},    \\
  &    &    &         &         &         & $-2.565$ \cite{dzuba-11}        \\
  &  2 & 1  & $2.707$& $2.834$& $1.607$& $2.682$ \cite{sahoo-11},           \\
  &    &    &         &         &         & $2.430$ \cite{dzuba-11}         \\
$^{139}$Ba$^+$&  
     3 & 3  & $-7.060$ & $-6.884$ & $-4.951$ & $-7.250$ \cite{sahoo-11}     \\
  &   &     &         &         &         & $-6.510$ \cite{dzuba-11}        \\
  &  2 & 3  & $6.888$& $7.386$& $6.096$& $7.389$ \cite{sahoo-11},           \\
  &   &     &         &         &         & $6.510$ \cite{dzuba-11}         \\
\hline
$^{225}$Ra$^+$&  
     2 & 1  & $-8.568$ & $-9.084$ & $-8.125$ & $-9.918$ \cite{sahoo-11},    \\
  &   &     &         &         &         & $-8.90$ \cite{dzuba-11}         \\
$^{223}$Ra$^+$&  
     3 & 2  & $-30.414$ & $-32.513$ & $-28.840$ & $-35.204$ \cite{sahoo-11},\\
  &   &     &         &         &         & $-31.65$ \cite{dzuba-11}        \\
  &  2 & 1  & $30.307$ & $32.286$ & $15.683$ & $30.525$ \cite{sahoo-11},    \\
  &   &     &         &         &         & $24.15$ \cite{dzuba-11}         \\
$^{229}$Ra$^+$&  
     2 & 3  & $-20.217$ & $-21.788$ & $1.137$ & $-16.297$ \cite{sahoo-11}   \\
  &  3 & 2  & $47.336$ & $50.917$ & $20.614$ & $-9.50$ \cite{dzuba-11}      \\
  &  2 & 2  & $-52.906$ & $-53.404$ & $-38.558$ & $-57.387$ \cite{sahoo-11},\\
  &   &     &         &         &         & $-49.00$ \cite{dzuba-11}        \\
\end{tabular}
\end{ruledtabular}
\label{e1pnc1}
\end{center}
\end{table}
                                                 
%
%
\begin{table}                                        
\begin{center}                                        
\caption{Component wise contribution from the coupled-cluster terms
         for $6\;^2S_{1/2} \rightarrow 7\;^2S_{1/2}$ transition in Cs,
         $6\;^2S_{1/2} \rightarrow 5\;^2D_{3/2}$ transition in
         Ba$^+$, and $7\;^2S_{1/2} \rightarrow 6\;^2D_{3/2}$
         transition in Ra$^+$.}
\begin{ruledtabular}                                                
\label{e1pnc_prcc}
\begin{tabular}{ccccccc}                                             
Atom & \multicolumn{2}{c}{Transition} & $D S^{(1)}_1$ & 
   ${S^{(1)}_1}^\dagger D$ & $D T^{(1)}_1$ & ${S^{(0)}}^\dagger D S^{(1)}_1$\\ 
\cline{1-5}                                                              
     & $F_f$  & $F_i$  &    &   &  $+$ c.c.  &       $+$ c.c.              \\
\hline                                                              
  $^{133}$Cs & 
  $3$ & $3$ & $-0.278$ & $4.198$ & $-0.005$ &$-1.533$               \\ 
 &$4$ & $4$ & $-0.317$ & $4.779$ & $-0.006$ &$-1.746$               \\ 
 &$4$ & $3$ & $0.764$ & $6.385$  & $-0.486$ &$-0.531$               \\ 
 &$3$ & $4$ & $0.657$ & $8.001$  & $-0.488$ &$-1.121$               \\ 
\hline                                                              
  $^{135}$Ba$^+$& 
  $3$ & $2$ & $-2.676$ & $0.590$ & $-0.647$ &$-0.301$               \\ 
 &$2$ & $1$ & $2.566$ & $-0.723$ & $0.603$ &$-0.276$                \\ 
  $^{139}$Ba$^+$&                   
  $3$ & $3$ & $-6.808$ & $1.729$ & $-1.620$ &$0.747$                \\ 
 &$2$ & $3$ & $6.784$ & $-1.496$ & $1.639$ &$-0.763$                \\ 
\end{tabular}                                                       
\end{ruledtabular}                                                  
\end{center}                                                        
\end{table}                                                        

{\em Results}.---For the calculations reported in the letter, we use Gaussian 
type orbitals 
generated with $V^{N-1}$ central potential. The $E1_{\rm PNC}^{\rm NSD}$ of 
Cs, Ba$^+$ and Ra$^+$ between various hyperfine states are given in 
Table. \ref{e1pnc_table}. There is a close match between our MBPT results and
results from similar works. 

 There are changes when the transition amplitudes are calculated
with PRCC. This can be attributed to the inclusion of higher order correlation
effects. However, it require a systematic series of calculations to examine 
the nature of the correlation effects from the higher order terms which
are subsumed in the PRCC calculations. The results of $^{229}$Ra$^+$ is a 
cause for concern, there is a large cancellation in the 
$F_i = 3 \rightarrow F_f=2$ transition amplitude.  However, for the other
two transitions of the same ion, the transition amplitudes are higher than 
Ba$^+$. In particular, the $F_i = 2 \rightarrow F_f=2$ transition amplitude
of $^{229}$Ra$^+$  is the largest among all the values and this is in 
agreement with the previous results. For the neutral atom Cs, the PRCC results
are larger than MBPT. This indicates, higher order correlation effects 
enahances $E1_{\rm PNC}^{\rm NSD}$. It is opposite in Ba$^+$ and Ra$^+$, the
PRCC results are lower than MBPT and indicates higher correlation effects
have suppression effect.

 To examine the impact of electron correlation in better detail, consider
the leading order (LO) and next to leading order (NLO) terms as listed in 
Table. \ref{e1pnc_prcc}. In the PRCC calculations, as given in 
Eq. (\ref{e1pnc_elec}), 
for Cs these are ${\pso}^{\dagger}D$ and $D\pso$, respectively. Here, the 
former represents $H_{\rm PNC}^{\rm NSD}$ perturbed $7\; ^2S_{1/2}$ and has 
larger opposite parity mixing as it is energetically closer to odd parity
states like $ 6\;^2P_{1/2}$. The same is  not true of $6\; ^2S_{1/2}$, which
is represented by $D\pso$.

 In the case of Ba$^+$ the LO and NLO are $D\pso$ and ${\pso}^{\dagger}D$, 
respectively. Although, not shown in Table. \ref{e1pnc_prcc} a similar pattern 
is observed in Ra$^+$. The sequence is opposite to Cs. Reason is,
the transitions in these ions are of 
$n\; ^2S_{1/2}\rightarrow n'\; ^2D_{3/2}$ type and 
matrix elements of 
$H_{\rm PNC}^{\rm NSD}$ involving $n'\; ^2D_{3/2}$ are negligible. Dominant 
contribution arises from the $sp$ matrix elements, which are large. So, the 
term $D\pso$, which represents $H_{\rm PNC}^{\rm NSD}$ perturbation of 
$n\; ^2S_{1/2}$ is the LO term of these ions.  The contribution from 
${\pso}^{\dagger}D$ is, however, non-zero as $n'\; ^2D_{3/2}$ acquires 
opposite parity mixing through electron correlation effects. It must be 
mentioned that, the Dirac-Fock contribution is the most dominant, however, in 
PRCC it is subsumed in the LO and NLO terms.  The terms which are second order 
in cluster operators, in Eq. (\ref{e1pnc_elec}), are non-zero but small. For 
comparison, the two dominant contributions from the second order term, 
${S^{(0)}}^\dagger D S^{(1)}_1$ and it's hermitian conjugate, is given in the 
Table. \ref{e1pnc_prcc}.

%
%
\begin{table}
\begin{center}
\caption{Reduced matrix element, $E1_{\rm PNC}^{\rm NSD}$, of the 
         $6\;^2S_{1/2} \rightarrow 5\;^2D_{5/2}$  and
         $7\;^2S_{1/2} \rightarrow 6\;^2D_{5/2}$ transitions between
         different hyperfine states  Ba$^+$ and Ra$^+$ respectively. 
         The values listed are in units of $iea_0\times 10^{-12}\mu'_W$.}
\begin{ruledtabular}
\label{e1pnc_d5b2}
\begin{tabular}{ccccccc}
Atom & \multicolumn{2}{c}{Transition} &\multicolumn{3}{c}{This work} 
                                      &Other works                     \\
\hline
  & $F_f$    &    $F_i$    &    DF    &    MBPT  &  PRCC   &           \\
\hline
$^{135}$Ba$^+$&$3$ & $2$&$0.003$& $0.098$& $0.227$&$0.041$ \cite{sahoo-11}\\
  &  $2$ & $1$  & $0.002$& $0.048$& $0.127$&      $-$        \\
$^{139}$Ba$^+$&$2$&$3$ & $0.003$&$0.125$& $0.235$& $0.043$ \cite{sahoo-11}\\
\hline
$^{223}$Ra$^+$&$3$&$2$& $-0.040$&$-0.605$& $1.262$& $-0.526$ \cite{sahoo-11}\\
$^{229}$Ra$^+$&$2$&$3$& $-0.019$&$-0.324$& $0.616$& $-0.256$ \cite{sahoo-11}\\
\end{tabular}
\end{ruledtabular}
\label{e1pnc1}
\end{center}
\end{table}

 We have also calculated the $E1_{\rm PNC}^{\rm NSD} $ of the 
         $6\;^2S_{1/2} \rightarrow 5\;^2D_{5/2}$  and
         $7\;^2S_{1/2} \rightarrow 6\;^2D_{5/2}$ transitions in 
Ba$^+$ and Ra$^+$, respectively, and the results are given in 
Table. \ref{e1pnc_d5b2}. The results from the PRCC are much larger than the 
MBPT results and this shows, without any ambiguity, electron correlation is 
the key to get meaningful results. This is on account of $d_{5/2}$ in 
the atomic $n\;^2D_{5/2} $ states, which are diffused and leads to larger
electron correlations.


{\em Conclusions}.---
The PRCC theory we have developed incorporates electron correlation effects 
arising from a class of diagrams to all order with 
a nuclear spin-dependent interaction as a perturbation. It is 
a suitable theory for precision calculations of atomic PNC arising from
$H_{\rm PNC}^{\rm NSD}$. With this method, it is possible
to incorporate electron correlation effects within the entire
configuration space obtained from a set of spin-orbitals.

{\em Acknowledgements}.---We thank B. K. Sahoo, S. Chattopadhyay, S. Gautam 
and S. A. Silotri for valuable discussions. The results presented in the paper 
are based on computations using the HPC cluster at Physical Research 
Laboratory, Ahmedabad.



\begin{thebibliography}{36}
   \bibitem{wood-97}
      C.~S.~Wood, et al. Science {\bf 275}, 1759 (1997).
   \bibitem{zeldovich-58}
      Y.~Zel'dovich,
      JETP {\bf 6}, 1184 (1958).
   \bibitem{fortson-93}
      N.~Fortson,
      Phys. Rev. Lett. {\bf 70}, 2383 (1993).
   \bibitem{tsigutkin-09}
       K.~Tsigutkin, {\em et al.},
      Phys. Rev. Lett. {\bf 103}, 071601 (2009).
   \bibitem{coester-58}
      F.~Coester,
      Nucl. Phys. {\bf 7}, 421 (1958).
   \bibitem{coester-60}
      F.~Coester and H. K\"ummel,
      Nucl. Phys. {\bf 17}, 477 (1960).
   \bibitem{hagen-08}
      G.~Hagen, T.~Papenbrock, D.~J.~Dean, and M.~Hjorth-Jensen,
      Phys. Rev. Lett. {\bf 101}, 092502 (2008).
   \bibitem{eliav-94}
      E.~Eliav, U.~Kaldor, and Y.~Ishikawa,
      Phys. Rev. A a{\bf 50}, 1121 (1994).
   \bibitem{nataraj-08}
      H. S. Nataraj, B. K. Sahoo, B. P. Das, and D. Mukherjee,
      Phys. Rev. Lett. {\bf 101}, 033002 (2008).
   \bibitem{pal-07}
      R.~Pal, {\em et al.},
      Phys. Rev. A {\bf 75}, 042515 (2007).
   \bibitem{isaev-04}
      T.~A.~Isaev, {\em et al.},
      Phys. Rev. A {\bf 69}, 030501(R) (2004).
   \bibitem{bishop-09}
      R.~F.~Bishop, P.~H.~Y.~Li, D.~J.~J.~Farnell, and C.~E.~Campbell,
      Phys. Rev. B {\bf 79}, 174405 (2009).
   \bibitem{sahoo-09}
      B.~K.~Sahoo, L.~W.~Wansbeek, K.~Jungmann, and R.~G.~E.~Timmermans,
      Phys. Rev. A {\bf 79}, 052512 (2009).
   \bibitem{thierfelder-09}
      C.~Thierfelder and P.~Schwerdtfeger,
      Phys. Rev. A {\bf 79}, 032512 (2009).
   \bibitem{sahoo-09a}
      B.~K.~Sahoo, B.~P.~Das, and D.~Mukherjee,
      Phys. Rev. A {\bf 79}, 052511 (2009).
   \bibitem{wansbeek-08}
      L.~W.~Wansbeek, {\em et al.},
      Phys. Rev. A {\bf 78}, 050501(R) (2008).
   \bibitem{porsev-10}
      S.~G.~Porsev, K.~Beloy, and A.~Derevianko,
      Phys. Rev. D {\bf 82}, 036008 (2010).
   \bibitem{pal-09}
      R.~Pal, D.~Jiang, M.~S.~Safronova, and U.~I.~Safronova,
      Phys. Rev. A {\bf 79}, 062505 (2009).
   \bibitem{sahoo-11}
      B.~K.~Sahoo, P.~Mandal and M.~Mukherjee,
      Phys. Rev. A {\bf 83 }, 030502 (2011). 
   \bibitem{latha-09}
      K.~V.~P.~Latha, D.~Angom, B.~P.~Das, and D.~Mukherjee,
      Phys. Rev. Lett. {\bf 103}, 083001 (2009).
   \bibitem{mani-10}
      B.~K.~Mani and D.~Angom,
      Phys. Rev. A {\bf 81}, 042514 (2010).
   \bibitem{mani-11}
      B.~K.~Mani and D.~Angom,
      Phys. Rev. A {\bf 83}, 012501 (2011).
   \bibitem{mani-09}
      B.~K.~Mani, K.~V.~P.~Latha, and D.~Angom,
      Phys. Rev. A {\bf 80}, 062505 (2009).
   \bibitem{sahoo-06}
      B.~K.~Sahoo, R.~Chaudhuri, B.~P.~Das, and D.~Mukherjee,
      Phys. Rev. Lett. {\bf 96}, 163003 (2006).
   \bibitem{lindgren-85}
      I.~Lindgren and J.~Morrison,
      {\it Atomic Many-Body Theory},
      edited by G.~Ecker, P.~Lambropoulos, and H.~Walther
      (Springer-Verlag, 1985).
   \bibitem{johnson-03}
      W.~R.~Johnson, M.~S.~Safronova, and U.~I.~Safronova, 
      Phys. Rev. A {\bf 67}, 062106 (2003).
   \bibitem{safronova-09}
      M.~S.~Safronova et al., 
      Nuclear Physics A {\bf 827}, 411c-413c (2009).
   \bibitem{dzuba-11}
      V.~A.~Dzuba, V.~V.~Flambaum,
      arXiv:1104.0086.
\end{thebibliography}
\end{document}